\documentclass[a4paper,11pt]{article}

\usepackage{jinstpub}

\usepackage{url}
\RequirePackage{color}

\usepackage{graphicx,epstopdf}
\usepackage{caption}
\usepackage{subcaption}
\emergencystretch=\maxdimen
\usepackage{url}
\usepackage{xfrac}
\usepackage[binary-units=true]{siunitx}
\DeclareSIUnit\torr{torr}
\usepackage{natbib}

\usepackage{tikz}
\usetikzlibrary{shapes,arrows}

\usepackage[english]{babel}

\title{Simulation of gain stability of THGEM gas-avalanche particle detectors}

\author[a,1,2]{P. M. M. Correia\note{Corresponding author.}\note{Equal contribution.},}
\author[b,2]{M. Pitt,}
\author[a]{C. D. R. Azevedo,}
\author[b]{A. Breskin,}
\author[b]{S. Bressler,}
\author[a]{C. A. B. Oliveira,}
\author[a]{A.L.M. Silva,}
\author[c]{R. Veenhof,}
\author[a]{and J. F. C. A. Veloso}

\affiliation[a]{I3N -  Physics Department\\
University of Aveiro\\
Campus Universit\' ario de Santiago 3810-193\\
Aveiro, Portugal}
\affiliation[b]{Dept. of Astrophysics and Particle Physics, Weizmann Institute of Science,\\
 P.O. Box 26, Rehovot 76100, Israel}
\affiliation[c]{CERN PH department,\\
CH-1211 Gen\`eve 23, Switzerland}

\emailAdd{pmcorreia@ua.pt}

\abstract{
Charging-up processes affecting gain stability in Thick Gas Electron Multipliers (THGEM) were studied with a dedicated simulation toolkit. Integrated with Garfield++, it provides an effective platform for systematic phenomenological studies of charging-up processes in MPGD detectors. We describe the simulation tool and the fine-tuning of the step-size required for the algorithm convergence, in relation to physical parameters.  Simulation results of gain stability over time in THGEM
detectors are presented, exploring the role of electrode-thickness and applied voltage on its evolution. The results show that the total amount of irradiated charge through electrode's hole needed for reaching gain stabilization is in the range of tens to hundreds of pC, depending on the detector geometry and operational voltage.  These results are in agreement with experimental observations presented previously.

}

\keywords{Detector modelling and simulations II (electric fields, charge transport, multiplication and induction, pulse formation, electron emission, etc); Micropattern gaseous detectors (MSGC, GEM, THGEM, RETHGEM, MHSP, MICROPIC, MICROMEGAS, InGrid, etc); Electron multipliers (gas)}

\begin{document}
\maketitle
\flushbottom

\section{Introduction}\label{sec:intro}

The time-evolution of the avalanche gain in Micro-Patterned Gaseous Detectors (MPGDs) has
been a topic of major interest in the last years. Gain variations with time
have been reported and are common in detectors containing dielectric materials
in contact with the gas medium where multiplication occurs; the effects have been noticed particularly in detectors based on hole-type MPGDs with insulating surfaces in contact with active gases, e.g. Gas Electron Multipliers (GEMs)\cite{Azmoun200611, industrialGEM, SAULI20162} and
Thick-GEMs (THGEMs)\cite{THGEM_operation_Ne_CH4, cortesi1, cortesi2,1748-0221-5-03-P03009,2006physics/0606162,Renous2017ogr,DallaTorre2015}.

Gain variation over time in GEM and THGEM-based detectors can arise from different processes, e.g. polarization of the insulator's volume under applied voltage, and charging-up of the insulator's surfaces exposed to the free drifting charges (electrons and ions) in the gas volume \cite{THGEM_operation_Ne_CH4}. It has been also observed that detector's environmental conditions, such as gas impurities and moisture, can affect gain transients\cite{Renous2017ogr}.

A simulation program has been developed for phenomenological studies of charging-up effects in GEM detectors\cite{Alfonsi20126, Correia2014}. The tool is based on the superposition principle according to which the field inside the holes is the sum of the field prior to any charge accumulating on the insulating surface and the field due to the charge accumulated on the insulators. This simulation tool predicted the time dependence of the gain in the GEM detectors with double-conical holes.

The superposition principle is also used to discuss experimental observation of gain variation over time \cite{Renous2017ogr,2017arXiv170900095L}. In this work, the simulation tools were further improved and extended to THGEM-based detectors. We have focused on the development of a new toolkit that allows for calculations of charging-up effects of the detector's insulator surfaces in a more efficient way. The polarization of the insulator volume and its influence on gain evolution were not studied in this document. In the current simulations, the detector was assumed to be biased prior to irradiation.

The present toolkit allows for the simulation of charging-up effects, based on the superposition principle of electric fields. This toolkit, which can be included to Garfield++\cite{garfieldpp}, can be extended to virtually all MPGDs in a straightforward way. It has been applied in this work to the study of gain evolution in THGEM detectors, under typical experimental conditions, and can be downloaded from\cite{github_chargingup}.

\section{The simulation toolkit}\label{sec:toolkit}

\subsection{Electric field calculations}\label{sec:algorithm}
To simulate charging-up effect in THGEMs, the
method for the charging-up calculations described in \cite{Correia2014}
was chosen.

The total field $\vec{E}_\text{tot}$ can be estimated using the superposition principle
\begin{equation}
\vec{E}_\text{tot}=\vec{E}_\text{uncharged}+\vec{E}_\text{charges}
\label{Field_superposition}
\end{equation}
Here $\vec{E}_\text{uncharged}$ is the electric field resulting from the voltage applied to the THGEM electrodes prior to any charging-up, $\vec{E}_\text{charges}$ is the field due to the charges accumulating in the insulator surfaces (electrons and ions).
This definition was also at the base of the analysis presented in \cite{Renous2017ogr}.
While $\vec{E}_\text{uncharged}$ remains constant, $\vec{E}_\text{charges}$ varies due to the continuous accumulation of charges.
Hence, the total field at equilibrium can be obtained in an iterative process, where in each iteration $\vec{E}_\text{charges}$ is modified according to the cumulative surface charge.

\begin{figure}[!ht]
\centering
\begin{subfigure}[c]{0.48\textwidth}
  \includegraphics[width=\linewidth]{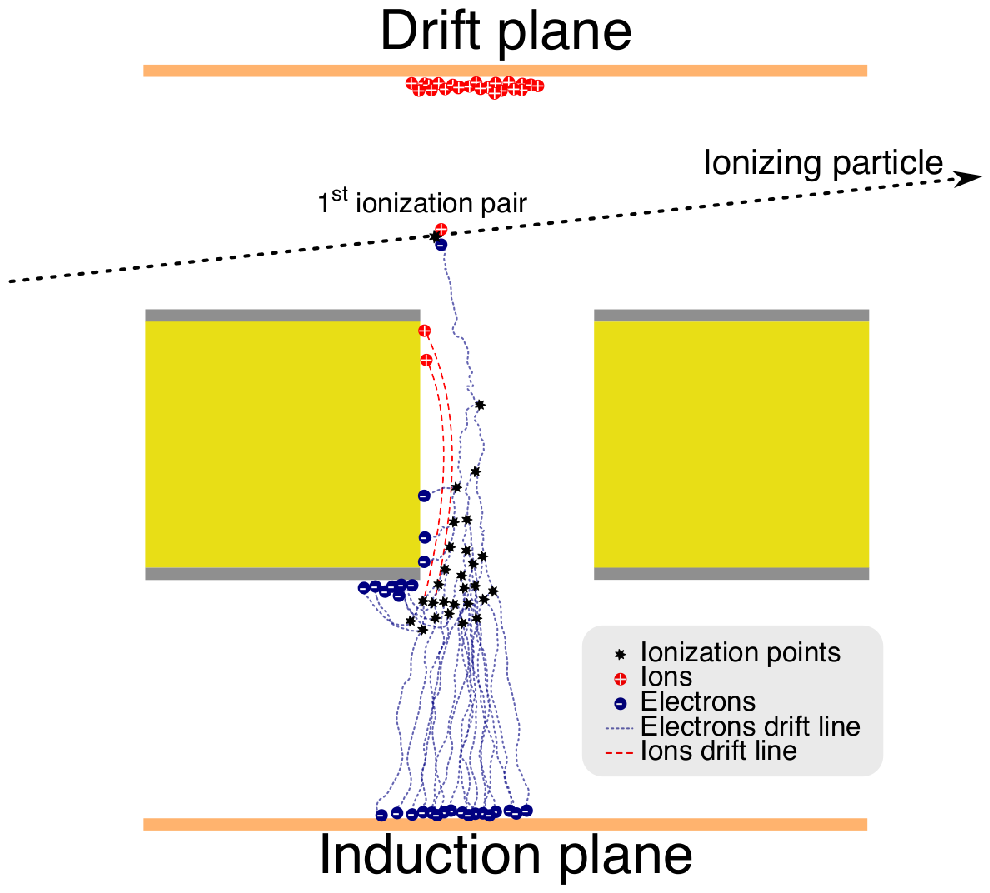}
  \caption{\label{fig:superposition_principle_avalanche}}
\end{subfigure}
\quad
\begin{subfigure}[c]{0.48\textwidth}
  \includegraphics[width=\linewidth]{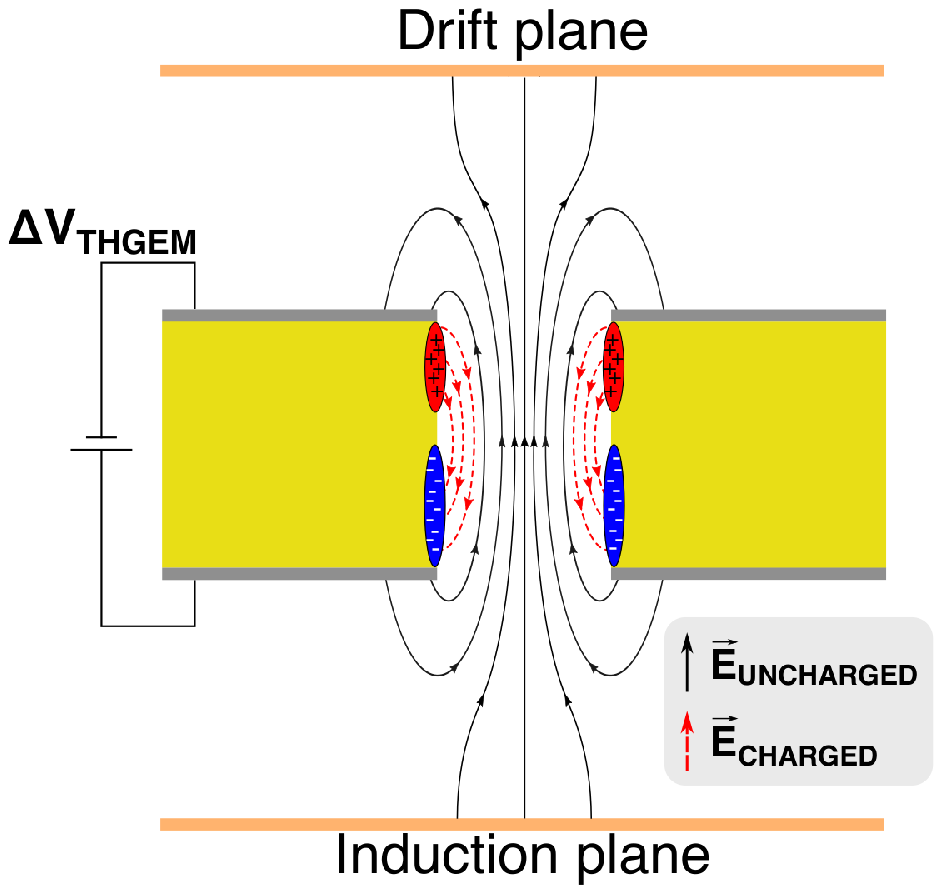}
    \caption{\label{fig:superposition_principle_simplified}}
\end{subfigure}

 \caption{(a) Schematic drawing of the charging-up process during an avalanche. (b) Charging-up superposition algorithm of the THGEM: the total electric field is a sum of $\vec{E}_\text{uncharged}$, calculated when voltages are applied on the electrodes and the detector was not exposed to ionizing radiation yet, superimposed with the electric field induced by accumulated charges on the insulator's surface of the detector $\vec{E}_\text{charges}$.
}
 \label{fig:superposition_principle}
\end{figure}

During the first iteration, that corresponds to
the simulation of the multiplication process in a THGEM before irradiation, i.e., prior to charge accumulation on the insulator surfaces, a set of primary avalanches is simulated. A fraction of the free charges (drifting electrons and ions) end up accumulating on the insulator, as depicted in Fig.~\ref{fig:superposition_principle_avalanche}. These charges will
modify the electric field by some amount, resulting in a new field map, $\vec{E}_\text{tot}$, that can be obtained from the superposition of $\vec{E}_\text{uncharged}$ and $\vec{E}_\text{charges}$, as depicted in Fig.~\ref{fig:superposition_principle_simplified}.

\subsection{Simulations steps}\label{sec:fieldmaps}
Previous simulation works related to charging-up effects in MPGDs relied on calculations
of the electric field in the region of interest and simulation of the avalanche and detector's gain.
The amount of charges produced on the avalanche that ended up trapped on the insulator surfaces was evaluated, a new electric-field map resulting from the addition of these charges was calculated, and the process was repeated, as shown in Fig.~\ref{fig:old_method_flowchart}.
Using this iterative method, the simulation results of the charging-up 		process were in agreement with experimental data obtained with GEM detectors \cite{Correia2014}.

The finite element method (FEM), as implemented in \textbf{ANSYS}\textregistered\footnote{www.ansys.com}, is used to calculate the electric potential in the gas volume, which then is used to infer the electric field at a given point of all the avalanche electrons' path\cite{COliveira2012JINST}.
Avalanche-size calculations and transport properties of the charges (electrons and ions) are calculated in Garfield++\cite{garfieldpp} (which interfaces with Magboltz \cite{magboltz,Biagi1999234}), that simulates collisions of electrons in a given gas mixture and detector geometry.

\tikzstyle{decision} = [diamond, draw, fill=blue!20,
    text width=4.5em, text badly centered, node distance=3cm, inner sep=0pt]
\tikzstyle{block} = [rectangle, draw, fill=blue!20,
    text width=5em, text centered, rounded corners, minimum height=4em]
\tikzstyle{line} = [draw, -latex']
\tikzstyle{cloud} = [draw, ellipse,fill=red!20, node distance=3cm,
    minimum height=2em]

    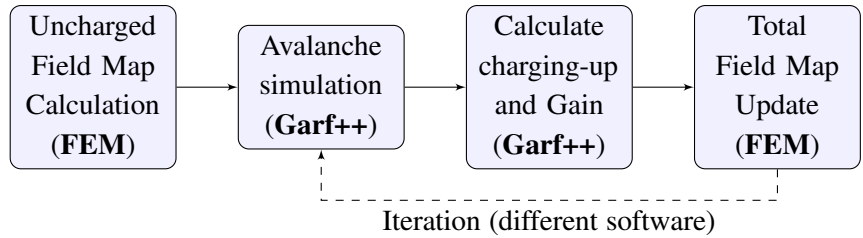
\begin{figure}[ht]

    \centering
\begin{tikzpicture}[node distance = 2cm, auto]
    \node [block,fill=blue!6] (uncharged) {Uncharged Field Map Calculation (\textbf{FEM})};
    \node [block,fill=blue!6, right of=uncharged, node distance=3cm] (avalsimu) {Avalanche simulation (\textbf{Garf++})};
    \node [block,fill=blue!6, right of=avalsimu, node distance=3cm] (charging) {Calculate charging-up and Gain (\textbf{Garf++})};
    \node [block,fill=blue!6, right of= charging, node distance=3cm] (iterate) {Total Field Map Update (\textbf{FEM})};

    \path [line] (uncharged) -- (avalsimu);
    \path [line] (avalsimu) -- (charging);
    \path [line] (charging) -- (iterate);
    \path [line,dashed] (iterate) --++ (0cm,-1.5cm) -| node [near start] {Iteration (different software)} (avalsimu);

\end{tikzpicture}
\caption{Flowchart of the previous simulation method, alternating from FEM back to Garfield++ after each iteration.\label{fig:old_method_flowchart}}
\end{figure}

The previous implementation of the simulation toolkit was CPU and time consuming. It was based on an iterative transition between the FEM software and Garfield++ after each iteration, as depicted in Fig.~\ref{fig:old_method_flowchart}. 
The update of the field-map, according to Garfield++ charge deposition results, required a new FEM calculation - to include the original electric field produced by the voltages applied in the detector as well as that arising from the accumulation of charges on the insulator surfaces. The simulations could take as long as several hundreds days of CPU time for a given gas mixture and amplification voltage. Even the use of parallelism tools (multiple processors, multithread, etc) can only reduce the computing time but not the algorithm efficiency, which sometimes even required manual intervention in the files between iterations.  

\begin{figure}[ht]

    \centering
\begin{tikzpicture}[node distance = 2cm, auto]
    \node [block,fill=blue!6] (uncharged) {Uncharged Field Map Calculation (\textbf{FEM})};
    \node [block,fill=blue!6, right of=uncharged, node distance=3cm] (avalsimu) {Avalanche simulation (\textbf{Garf++})};
    \node [block,fill=blue!6, right of=avalsimu, node distance=3cm] (charging) {Calculate charging-up and Gain (\textbf{Garf++})};
    \node [block,fill=blue!6, right of= charging, node distance=3cm] (iterate) {Total Field Map Update (\textbf{Garf++})};

    \path [line] (uncharged) -- (avalsimu);
    \path [line] (avalsimu) -- (charging);
    \path [line] (charging) -- (iterate);
    \path [line,dashed] (iterate) --++ (0cm,-1.5cm) -| node [near start] {Iteration (superposition calculations)} (avalsimu);

\end{tikzpicture}
\caption{Flowchart of the superposition method. Compared with previous flowchart (Fig.~\protect\ref{fig:old_method_flowchart}), FEM is used only at the beginning of the simulation. All further calculations are performed in Garfield++.\label{fig:new_method_flowchart}}
\end{figure}
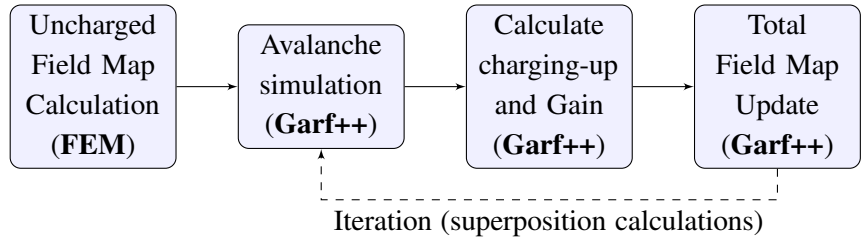

Hence, in this work we have elaborated a faster solution based in a different algorithm. While in the previous method, the new filed-maps between each iterations were calculated in the FEM software, forcing a transition between the FEM software and Garfield++ during after each iteration, in the method now described, several field-maps, previously calculated in the FEM software, are imported to Garfield++, and  further iterative calculations are performed without running new instances of the FEM software, speeding up the simulations. A schematic description is given in Fig.~\ref{fig:new_method_flowchart}.

A new $\vec{E}_\text{charges}$ is calculated at the end of each iteration and is used to updated $\vec{E}_\text{tot}$.
The next iterations takes into consideration the updated
$\vec{E}_\text{tot}$ of the previous iterations. The different electric-field during iterations changes the avalanche process, which influences on the
measured gain. Assuming a certain irradiation rate, a time-evolution of the gain can therefore be simulated. A flowchart of the superposition algorithm is depicted in Fig.~\ref{fig:new_method_flowchart}.

\begin{figure}[!ht]
    \centering
        \includegraphics[width=0.4\textwidth]{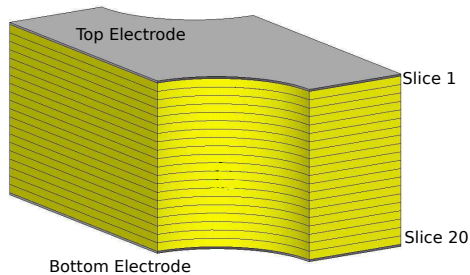}
    \caption{The THGEM cell (without a hole-rim) employed in the simulations. The insulator surface is divided into 20 slices along the THGEM hole.
}\label{fig:thgem_cross_section_and_cell}
\end{figure}

For a THGEM, the following procedure is applied for computing the field maps of interest:
\begin{itemize}
  \item The THGEM insulator surface is divided into 20 slices as shown in Fig.~\ref{fig:thgem_cross_section_and_cell}.
  \item The $\vec{E}_\text{uncharged}$ is calculated from the voltages applied;
  \item For each slice, a field map, considering only the electric field due to an
unitary charge distributed uniformly over the exposed surface of the slice, is calculated - defined here as $\vec{E}_\text{slices}$.
\end{itemize}

Once these field maps are available, the Garfield++ simulation can start.
To perform the simulation, Garfield++ imports the
coordinates of the nodes and the associated electric potential as a list.
In order to calculate $\vec{E}_\text{charges}$, an avalanche is calculated, and the position of all the electrons or ions deposited on an insulating surface is stored. The total number of accumulated charges
$N = N_\text{electrons} - N_\text{ions}$ (the number of accumulated electrons minus that of accumulated ions) is assigned to the corresponding slice. Since the expected variation of the field from $N$ charges on a given slice for a single avalanche is very small, for fast algorithm convergence, the total number of accumulated charges attached to a single slice is usually multiplied by a constant value $s$ (\textit{step-size}). Therefore, in the case where $N$ charges have stopped on the surface of a slice $j$ (without other extra charges accumulated in the remaining surfaces), we need to calculate, for each node $i$, the
electric potential $V$ for that node which is given by Eq.~\ref{eq:superposition}:
\begin{equation}
  V(charges,i)=V(uncharged,i)+N\times s\times V(j,i)
  \label{eq:superposition}
  \end{equation}
where $V(j,i)$ is the electric potential on node $i$ due to the presence
of a unitary positive charge in the surface of slice $j$, calculated from the $\vec{E}_\text{slices}$ described previously. Depending on the accumulated charges, $N$ can either be positive (more ions) or negative (more electrons).

Once $\vec{E}_\text{charges}$ is calculated, the next iteration is launched, where the total electric field is given as a sum of $\vec{E}_\text{charges}$ and $\vec{E}_\text{uncharged}$ (Eq.~\ref{Field_superposition}). In the following iteration, a new $\vec{E}_\text{charges}$ is calculated according to a new amount of accumulated charges, and this procedure is repeated until the total amount of accumulated charges $N$ after a given number of iterations is equal to zero - $\vec{E}_\text{tot}$ became constant with the increasing number of iterations.

\section{Application of the method to THGEMs}\label{sec:Simulations}
\subsection{Simulation setup}\label{sec:simulation_setup}
With the method described in the previous section, we are able to
simulate the charging-up effect using
only Garfield++, even though more than one field map has to be introduced as input at the beginning of the simulation.

The THGEM-detector geometry simulated in this work has an electrode with an hexagonal pattern of hole-pitch \textit{a} = \SI{1}{\milli\meter}, hole diameter \textit{d} = \SI{0.5}{\milli\meter}, with no rim,
and FR4 substrate having a thickness \textit{t} = \SI{0.4}{\milli\meter}, \SI{0.8}{\milli\meter} and \SI{1.2}{\milli\meter}.
Electric fields of 0.2 and \SI{0.5}{\kilo\volt\per\centi\meter} were set
across the drift and induction gaps respectively.

Voltages $\Delta V_{\text{THGEM}}$ were applied across the THGEM electrode.
Several gas mixtures were used in the simulations. In most of the cases
Ne and Ne/CH$_{4}$(5\%), however to obtain similar electron multiplication at higher voltages, the simulation work was extended to Ar-based gas mixtures: Ar, Ar/CH$_4$ (5\%)
and Ar/CO$_2$(5,7\%).
Since some of these are believed to be Penning-mixtures \cite{1952PhRv.88.417J}, Penning factors of
0.40 for Ne/CH$_{4}$(5\%), 0.18 for Ar/CH$_4$ (5\%),
0.35 for Ar/CO$_2$(5\%) and
0.4 for Ar/CO$_2$(7\%)\cite{AroucaTHGEMGAIN,Ozkan2010JINST,Sahin2014104} were used.
Standard room-temperature conditions were considered (\textit{T}=\SI{293}{\kelvin}
and \textit{P}=\SI{1}{\bar}).
A constant irradiation rate of \SI{10}{\hertz\per\milli\metre\squared}, \SI{8}{\kilo\electronvolt} x-rays,  was assumed. In our simulations, a random location for primary charges originated by the incoming x-rays is assumed and the charges are drifted towards the THGEM holes from different positions. 

At each iteration, the gain for a single-electron avalanche multiplication was calculated. Electron multiplication is a stochastic process, and the gain is determined as the average number of electrons generated during the electron avalanche process. At each iteration a given number of avalanches between 100 and 1000 was simulated (denoted in the text as $n_{\mathrm{AV}}$). In all simulations, the statistical error on the mean gain of the detector was taken to be
\begin{equation}
\frac{\sigma\mathrm{(G)}}{\mathrm{G}}=\sqrt{\frac{1}{n_{\mathrm{AV}}}}
\end{equation}
In each avalanche, the position of all the electrons or ions deposited on an insulating surface is stored, and assigned to the corresponding slice. Electrons that do not end up on the insulator are either collected on
the THGEM bottom electrode, on the induction plane, while non-trapped ions end up on the top THGEM electrode
or on the drift plane. Some of the electrons are attached to the gas molecules (CO$_2$ or CH$_4$) and are not
taken into account in the gain calculation.

\subsection{Physical interpretation}\label{sec:interpretation}

The number of simulated iterations ($n_\mathrm{iter}$) can be translated to a physical irradiation time ($t$) as follows: given the number of primary charges induced in the gas by the ionizing event\footnote{\SI{8}{\kilo\electronvolt} X-rays produce \textasciitilde222 or \textasciitilde300 primary electrons on average in Ne-based or Ar-based gas mixtures\cite{Sauli:117989}, respectively, per interaction} ($n_\mathrm{p}$), the irradiation rate ($R$) and the \textit{step-size} ($s$) can be related to a physical irradiation time:

\begin{equation}
  t[\mathrm{sec}] = \frac{s}{n_\mathrm{p}\times R\ [\mathrm{Hz}]} \times n_\mathrm{iter}
  \label{eq:step_time_equation}
\end{equation}

\begin{figure}[!ht]
    \centering
    \begin{subfigure}[t]{0.45\textwidth}
        \centering
        \includegraphics[width=\textwidth]{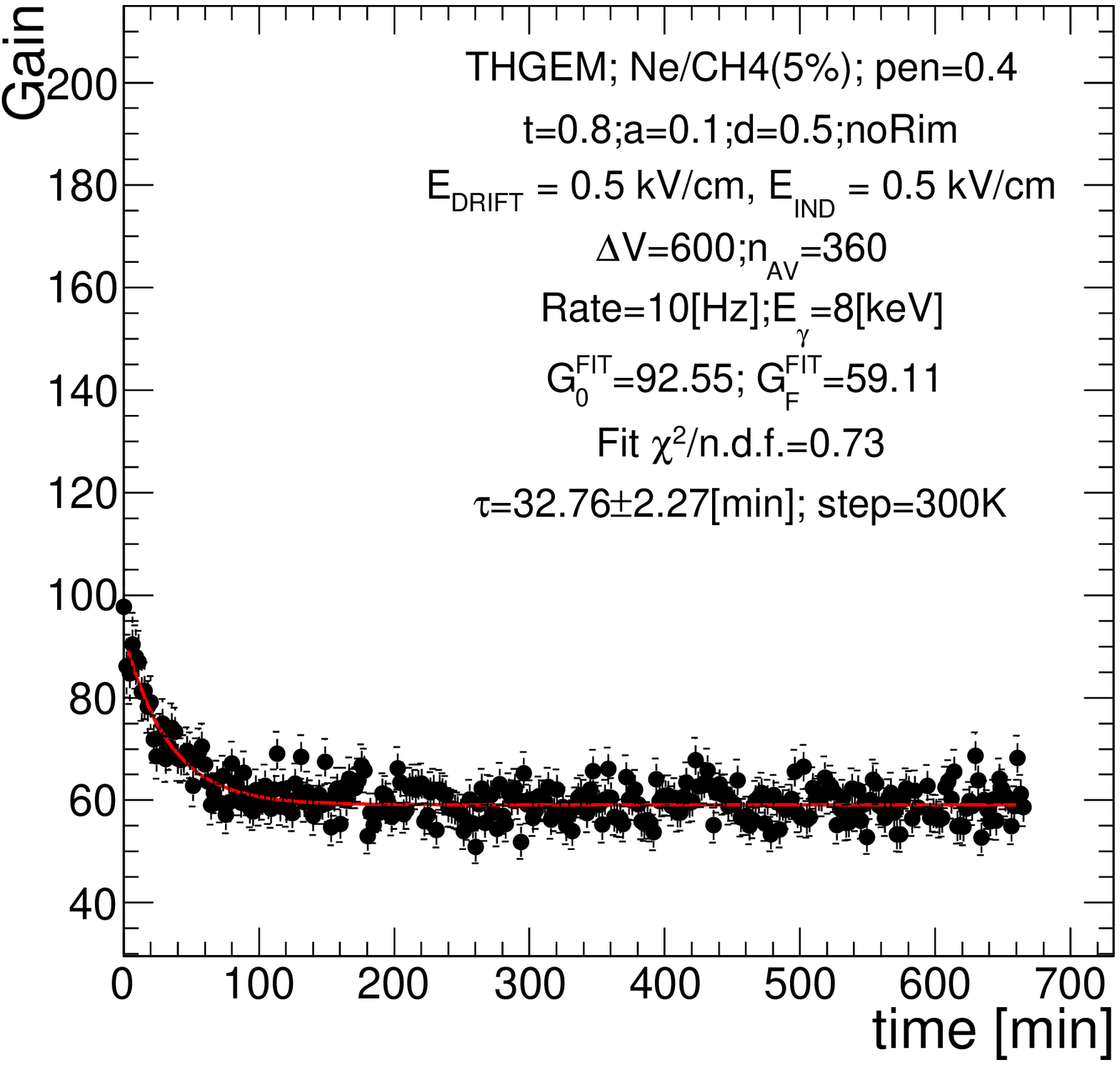}
        \caption{}
        \label{fig:gain_small_component_a}
    \end{subfigure}%
    ~
    \begin{subfigure}[t]{0.45\textwidth}
        \centering
        \includegraphics[width=\textwidth]{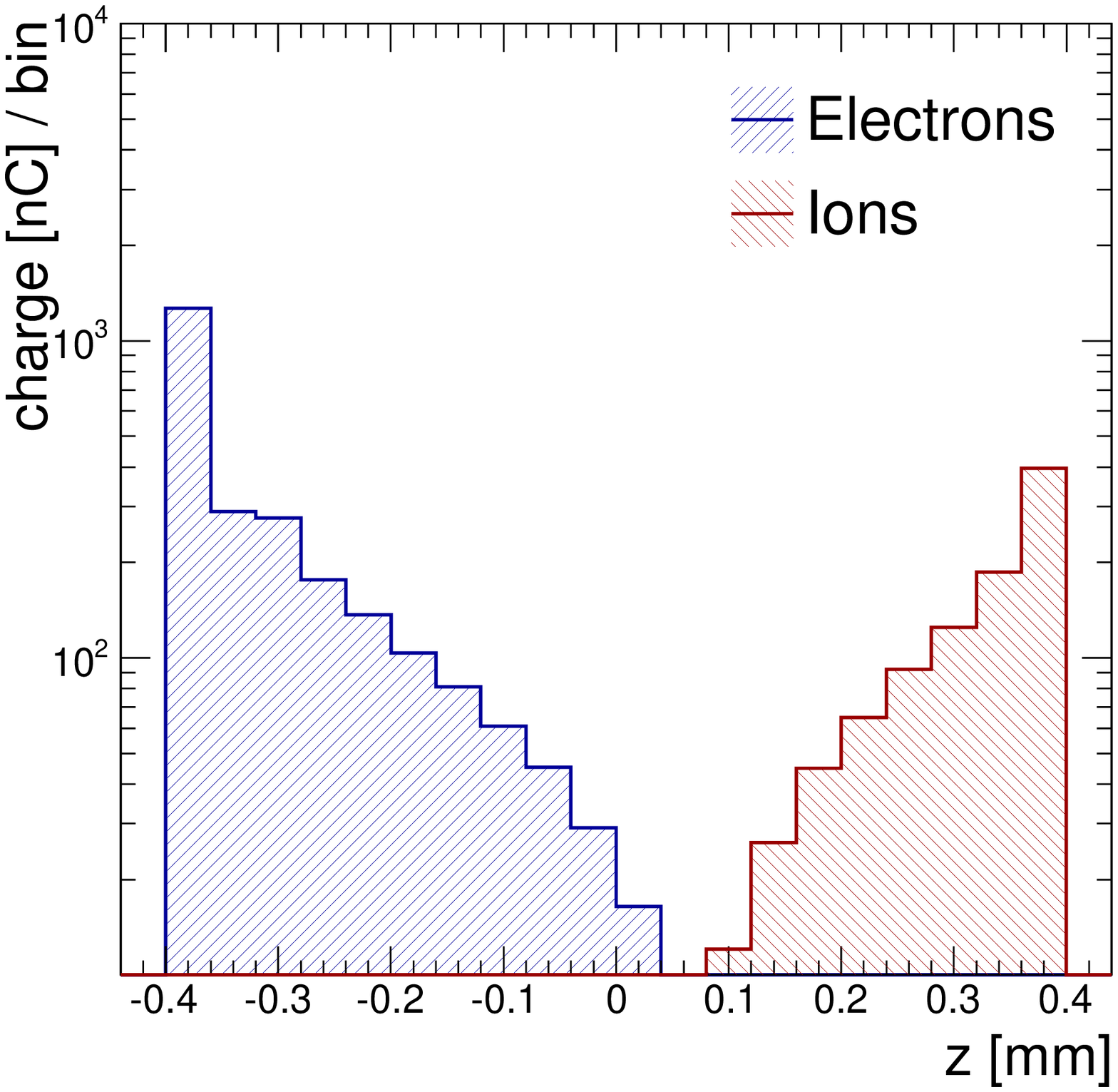}
        \caption{}
        \label{fig:gain_small_component_b}
    \end{subfigure}
    \caption{Simulated gain evolution in the THGEM as function of time (a) and the distribution of the accumulated charges onto the insulating hole's walls, along the hole's z-axis (b), in Ne/CH$_{4}$(5\%). The exponential fit in the left figure relies on Eq.~\protect\ref{eq:gain_time_equation}. The fitted parameters are shown in the inner caption.}
    \label{fig:gain_small_component}
\end{figure}

An example of a simulation result of the gain evolution in a THGEM detector, in Ne/CH$_{4}$(5\%) gas
mixture with the parameters: \textit{t}=\SI{0.8}{\milli\meter}, \textit{a}=\SI{1}{\milli\meter}, 
\textit{d}=\SI{0.5}{\milli\meter}, and no-rim, is shown in Fig.~\ref{fig:gain_small_component}.

Gain evolution curves over time were fitted using:
\begin{equation}
G(t)=G_0 - \Delta G \big(1 - e^{-t/\tau}\big)
\label{eq:gain_time_equation}
\end{equation}

where $G(t)$ is the gain at a given instant, $G_0$ is the initial
gain, $\tau$ is the relaxation time constant (or \textit{characteristic time}),
and $\Delta G$ is the gain variation.
The final gain, $G_\mathrm{F}$, can be calculated in terms of $G_0$ and $\Delta G$:
 \begin{equation}
G_\mathrm{F} = G(+\infty) = G_0 - \Delta G
\label{eq:init_gain}
\end{equation}

The quality of the fit of Eq.~\ref{eq:gain_time_equation} indicates upon a good agreement
($\chi^2/\mathrm{n.d.f.}=0.71$) to the simulation (n.d.f., number of degrees of freedom, being the number of iterations). The exponential model is driven by the assumption that the amount of the accumulated charges is proportionally decreasing with the amount of charges attached to the insulating surfaces. In Fig.~\ref{fig:gain_small_component}, after about \textasciitilde\SI{100}{\minute}, the gain reaches a plateau, indicating that the
ongoing charge accumulation is no longer capable of modifying the electric field
in a way that could affect the gain because the electric fields due to new incoming electrons and ions cancel each other. 

Using the formalism above, one can define a \textit{characteristic charge}, $Q_\mathrm{tot}$ - the total
number of charges, which pass through the THGEM holes during the relaxation period ($\tau$):

\begin{equation}
Q_\mathrm{tot} =n_\mathrm{p}\times R \times \int_{0}^{\tau}G(t)dt=n_\mathrm{p}\times R \times \tau \times
\big(G_0 - \frac{\Delta G}{e}\big)
\label{eq:Qtot_equation}
\end{equation}

Replacing the \textit{characteristic time} by \textit{characteristic number of iteration} ($\tau_\mathrm{iter}$), using Eq.~\ref{eq:gain_time_equation}, one can rewrite Eq.~\ref{eq:Qtot_equation} in terms of the \textit{step-size} and $\tau_\mathrm{iter}$:

\begin{equation}
Q_\mathrm{tot} =s\times \tau_\mathrm{iter} \times \big(G_0 - \frac{\Delta G}{e}\big)
\label{eq:Qtot_equation2}
\end{equation}

\begin{figure}[!ht]
    \centering
    \begin{subfigure}[t]{0.3\textwidth}
        \centering
        \includegraphics[width=\textwidth]{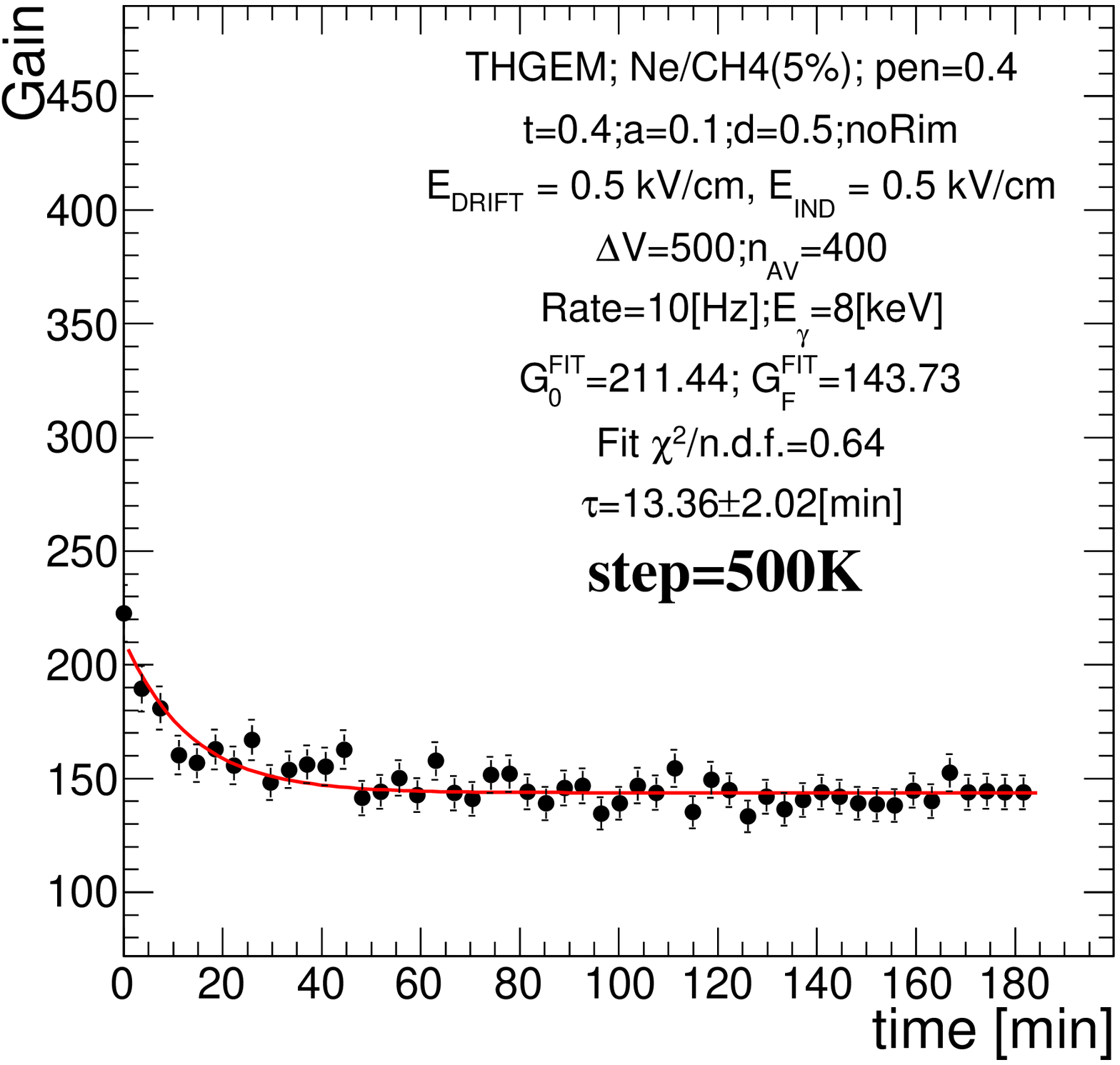}
        \caption{}
        \label{fig:gain_evo_steps_500K}
    \end{subfigure}%
    ~
    \begin{subfigure}[t]{0.3\textwidth}
        \centering
        \includegraphics[width=\textwidth]{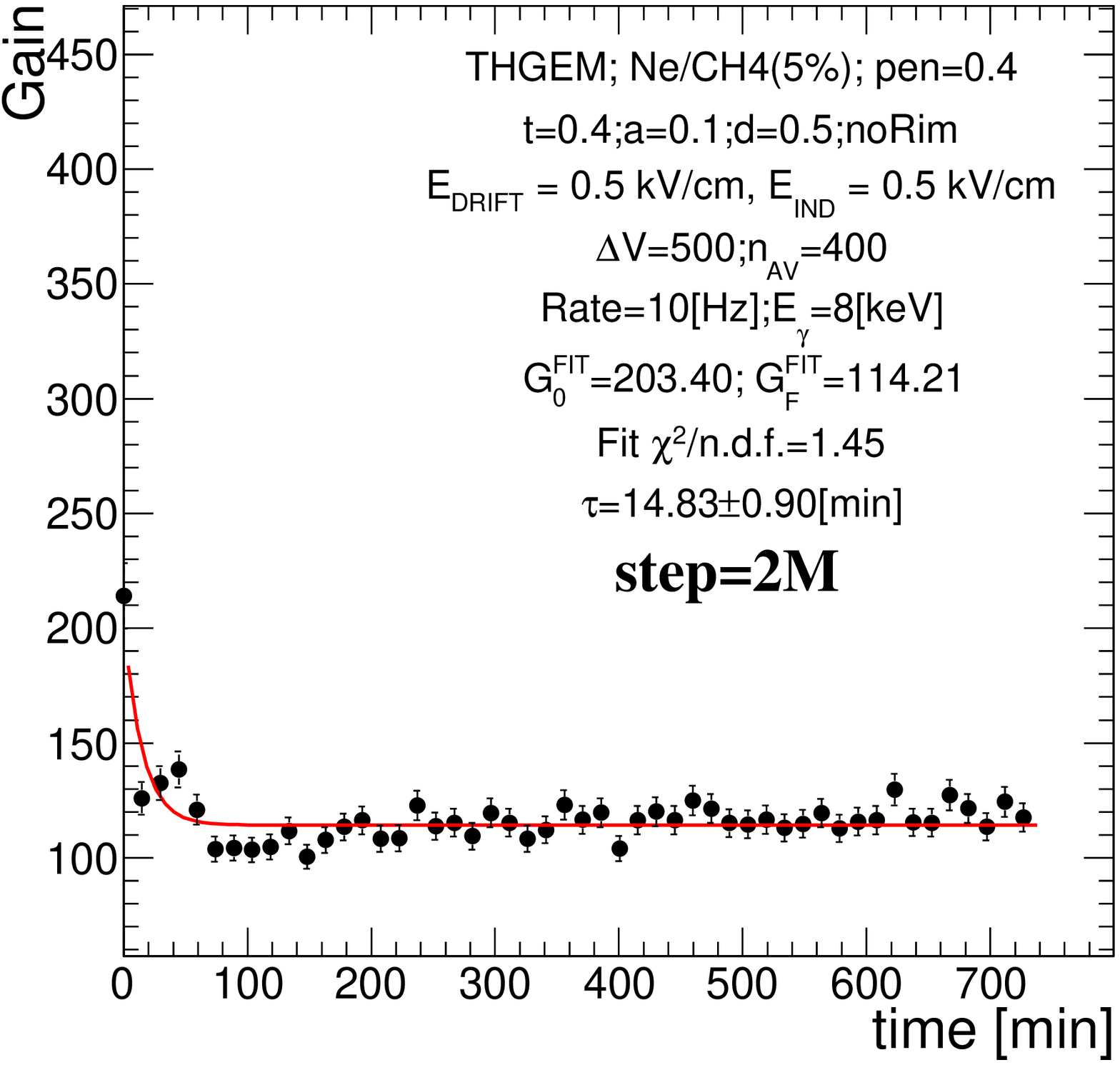}
        \caption{}
        \label{fig:gain_evo_steps_10M}
    \end{subfigure}
    ~
    \begin{subfigure}[t]{0.3\textwidth}
        \centering
        \includegraphics[width=\textwidth]{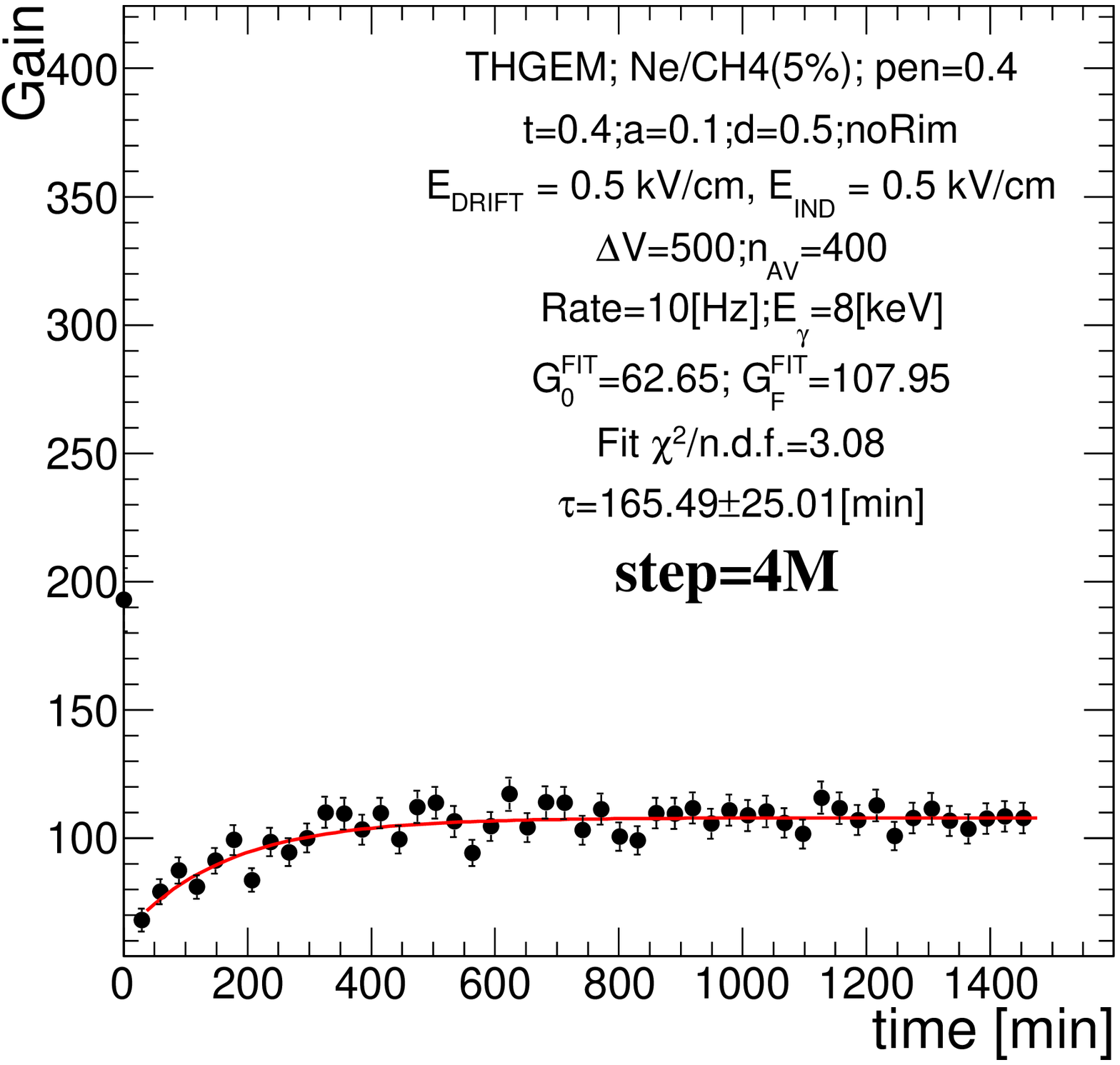}
        \caption{}
        \label{fig:gain_evo_steps_20M}
    \end{subfigure}
    \caption{Simulation results of gain stabilization for different step-size values. (a) Gradual decrease of gain due to moderate step-size, (b) the electrode fully charges up after the first iteration, (c) too large step-size.}
\label{fig:gain_evo_steps}
\end{figure}

The choice of the \textit{step-size s} is an important parameter for simulation convergence. An example of simulation results of THGEM detector, in Ne/CH$_{4}$(5\%) gas mixture with the parameters: \textit{t}=\SI{0.4}{\milli\meter}, \textit{a}=\SI{1}{\milli\meter}, \textit{d}=\SI{0.5}{\milli\meter}, and no-rim,
 with different \textit{step-size} values (from \num{0.5e6} to \num{40e6}) is shown in Fig.~\ref{fig:gain_evo_steps}. The behaviour of the gain $G(t)$ should remain
unchanged, regardless the \textit{step-size}. In Fig.~\ref{fig:gain_evo_steps_500K}
and~\ref{fig:gain_evo_steps_10M} the fitted $\tau$ is similar, while in Fig.~\ref{fig:gain_evo_steps_10M}
and~\ref{fig:gain_evo_steps_20M} clearly show that higher $step-size$ lead to different
trends.

\subsection{Convergence rules}\label{sec:Convergence}
\begin{figure}[!ht]
    \centering
    \begin{subfigure}[t]{0.45\textwidth}
        \centering
        \includegraphics[width=\textwidth]{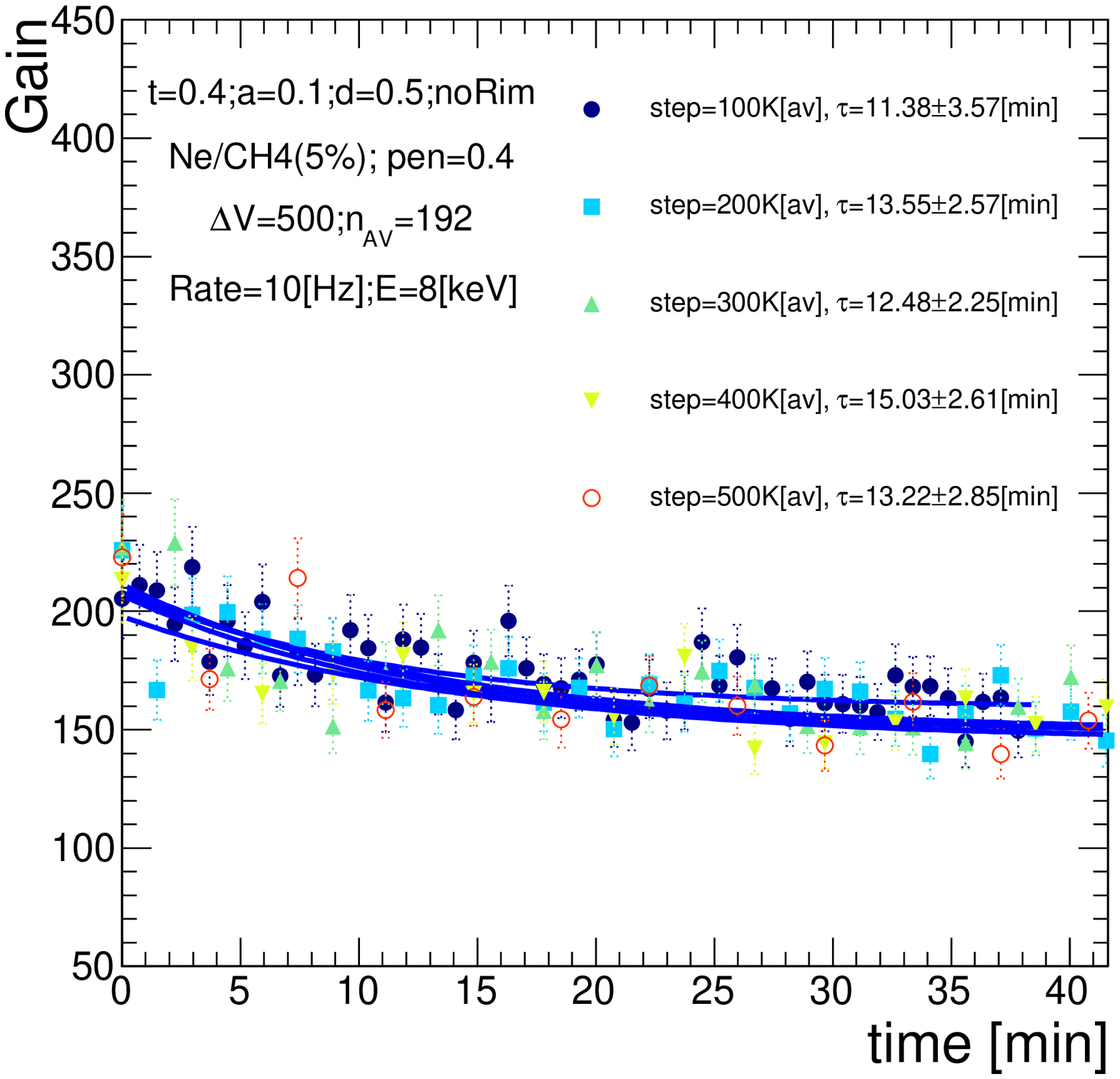}
        \caption{}
        \label{fig:gain_vs_geometry_vs_voltage_a}
    \end{subfigure}%
    ~
    \begin{subfigure}[t]{0.45\textwidth}
        \centering
        \includegraphics[width=\textwidth]{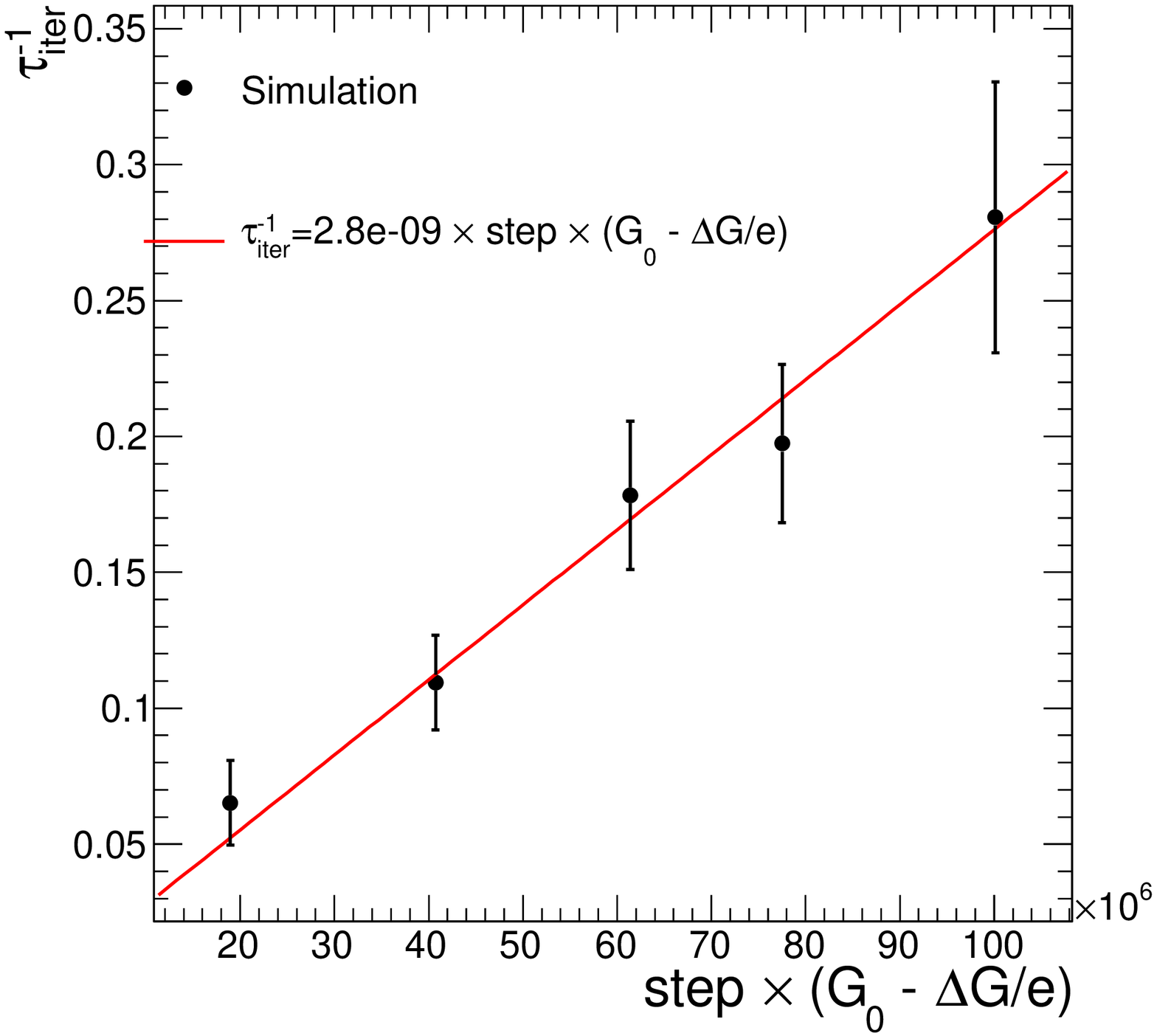}
        \caption{}
        \label{fig:StepFit_t04_a10_d05_h0}
    \end{subfigure}
    \caption{(a) Simulated gain stabilization in the THGEM detector, calculated in the time domain, the solid blue lines are the fitted exponential curves (from Eq.~\protect\ref{eq:gain_time_equation}). (b) The linear relation between the characteristic times in iteration domain vs the \textit{step-size}.}
\label{fig:one_over_tau_iter_plot}
\end{figure}

Since the \textit{step-size} should not affect the gain over time, a plot of the gain stabilization over time, simulated for different \textit{step-size} values from \num{0.1e6} to \num{0.5e6}, for the same detector geometry, is shown in the Fig.~\ref{fig:gain_vs_geometry_vs_voltage_a}, only for those \textit{step-size} values that do not vary the gain trend. According to the results obtained in Fig.~\ref{fig:gain_evo_steps_500K}, for a step size of \num{0.5e6} the gain stabilizes after about 3-4 iterations, therefore the step-size was reduces by a factor of 5 to inspect the gain stabilization over 50 simulated iterations.  From these curves, the average fitted $\tau$
is \SI{13(2)}{\minute}. The plot of $1/\tau_{\mathrm{iter}}$ as a function of
$s\times (G_0 - \frac{\Delta G}{e})$, depicted in the
Fig.~\ref{fig:StepFit_t04_a10_d05_h0}, allows to calculate the value of
$Q_{\mathrm{tot}}$ = \num{3.57e8} charges ($\approx$ \SI{57}{\pico\coulomb}).

\section{Results}
\subsection{Voltage effect on gain stabilization}\label{sec:voltageEffect}

Simulations were performed at different THGEM operation voltages, in the range $\Delta V_\mathrm{THGEM}$ = 500 - \SI{1100}{\volt}, aiming at investigating their influence on gain evolution.
To compare the gain evolution for a broader range of $\Delta V_\mathrm{THGEM}$ values, the simulation was performed for different gas mixtures (with the highest operation voltages in Ar mixtures).

\begin{figure}[!ht]
\centering
 \includegraphics[width=0.7\textwidth]{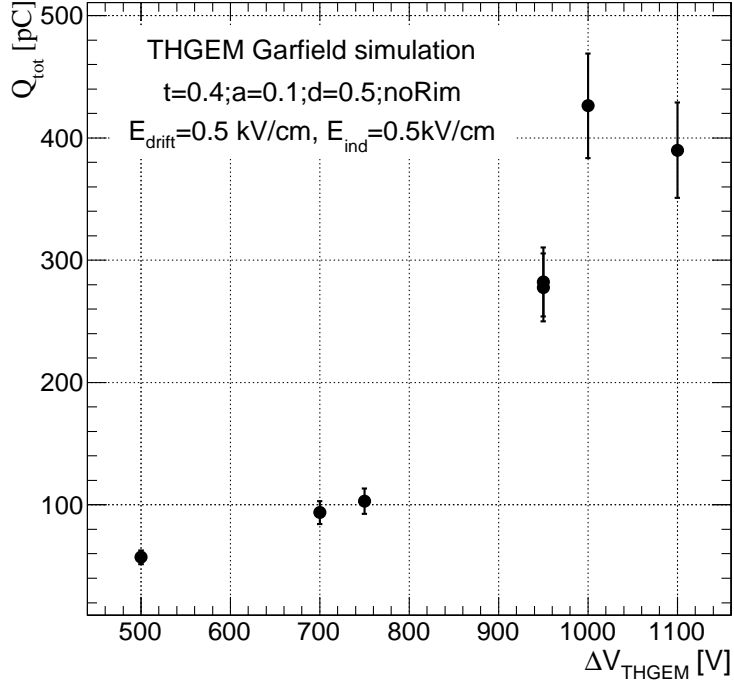}
 \caption{Simulated $Q_{tot}$ for different applied voltages across the THGEM. The $Q_{tot}$ is calculated according to Eq.~\protect\ref{eq:Qtot_equation2}. The error bars indicate the spread of calculated $Q_{tot}$ values among few tens of simulations with varying \textit{step-size}.}
 \label{fig:voltageEffect}
\end{figure}

In Fig.~\ref{fig:voltageEffect}, the calculated $Q_{tot}$ value (obtained from Eq.~\ref{eq:Qtot_equation2}) increases with voltage, indicating that the higher the voltage, the higher the irradiation rate (or time) is needed to reach gain stabilization.

\subsection{Electrode-thickness effect on the gain stabilization}\label{sec:thicknessEffect}

Gain stabilization was simulated for THGEM electrodes of different thickness,
in Ne/CH$_{4}$(5\%) and Ar/CO$_{2}$(5\%) (Ar-based mixtures have about two-fold higher operational voltages, and lower transverse electron diffusion for electric fields below \SI{10}{\kilo\volt\per\cm} \cite{magboltz}).
The $Q_{tot}$ value calculated from the charging-up of THGEM hole walls and the fraction of the avalanche charges attached to the THGEM walls for an uncharged electrode are shown in Table~\ref{tab:thicknessEffect}.

\begin{table}[!ht]
 \begin{center}
   \begin{tabular}{cccccc}
     Gas & Thickness & $\Delta V_\mathrm{THGEM}$ & $G_0$ & $Q_{tot}$ [\SI{}{\pico\coulomb}]  & Fraction [\%] \\
     \hline\hline
Ne/CH$_{4}$(5\%) & 0.4 & 450 & 80 & 48.11 & 23.4$\pm$1.8 \\
Ne/CH$_{4}$(5\%) & 0.4 & 500 & 220 & 57.14 & 23.1$\pm$1.8 \\
Ne/CH$_{4}$(5\%) & 0.8 & 600 & 110 & 34.25 & 31.9$\pm$0.8 \\
Ne/CH$_{4}$(5\%) & 1.2 & 750 & 130 & 30.43 & 38.0$\pm$0.7 \\
Ar/CO$_{2}$(5\%) & 0.4 & 950 & 120 & 277.75 & 18.3$\pm$0.3 \\
Ar/CO$_{2}$(5\%) & 0.8 & 1300 & 130 & 130.61 & 22.9$\pm$1.4 \\
     \hline
   \end{tabular}
   \caption{\label{tab:thicknessEffect}
     The simulated $Q_{tot}$ for different electrode thickness and operation voltages and the fraction
      (in \%) of the avalanche charge for uncharged THGEM electrode that accumulates on the THGEM-hole walls.}
\end{center}
\end{table}

An increase of the $Q_{tot}$ is observed with the decrease of the electrode thickness; it is found to be more pronounced than the dependence on the THGEM voltages or electron transport parameters, e.g. for thicker structures, and higher applied voltages, a lower total-charge value would be needed for modifying the gain. For thicker electrodes, the fraction of accumulated charges on the THGEM holes is larger, due to a broader transverse expansion of the avalanche.

\newpage
\section{Conclusions}\label{sec:conclusions}
The present study introduced a new toolkit for charging-up calculations in MPGDs that can be included in a future version of Garfield++; it is expected to ease simulations of phenomena occurring in gas-avalanche detectors. A case-study is shown, where the new toolkit was used to model the gain stabilization due to charging-up effects in THGEM detectors; a quantitative study was conducted to characterize various aspects of these effects. The gain stabilization typically occurs in a few minutes to a few tens of minutes, depending on several parameters of the THGEM-electrodes. Higher voltages applied across the THGEM-electrode require a larger amount of charge to modify the electric field within the detector holes, resulting in longer gain-stabilization characteristic times. On the other hand, in thicker electrodes, the avalanche is broader due to the expansion of the avalanche-electrons within the holes, and the charging-up of the electrode's insulating surface occurs faster. Most of the simulation studies were carried at relatively low operation voltages (low gains). Differently to GEM detectors with single-conical or double-conical geometry, where the gain stabilization exhibits an increase over time, in THGEM electrodes with cylindrical holes, the effect is a decrease of the detector gain due to a different charge distribution along the hole's surface.
A typical characteristic charging-up time in THGEM electrodes is found to be equivalent to a total irradiated charge of hundreds of \SI{}{\pico\coulomb} similarly to the values reported for the GEM electrodes with double-conical geometry.
 The simulation results of the initial decrease of the gain are found to be in a good agreement with previously reported experimental data \cite{cortesi1,cortesi2}. Further experimental studies are in course to validate our charging-up model.\\

\acknowledgments\
This research was supported in part by the I-CORE Program of the Israel Planning and Budgeting Committee, the Nella and Leon Benoziyo Center for High Energy Physics at the Weizmann Institute of Science\\
C.D.R.~Azevedo and P.M.M.Correia were supported by FCT (Lisbon) grants SFRH/BPD/79163/2011 and  PD/BD/52330/2013 and by I3N laboratory, funded by UID/CTM/50025/2013.

\bibliographystyle{JHEP}
\bibliography{charging}

\providecommand{\href}[2]{#2}\begingroup\raggedright\begin{thebibliography}{10}

\bibitem{Azmoun200611}
B.~Azmoun, W.~Anderson, D.~Crary, J.~Durham, T.~Hemmick, J.~Kamin et~al.,
  \emph{A study of gain stability and charging effects in \mbox{GEM} foils}, .

\bibitem{industrialGEM}
F.~Simon, B.~Azmoun, U.~Becker, L.~Burns, D.~Crary, K.~Kearney et~al.,
  \emph{Development of tracking detectors with industrially produced gem
  foils}, \href{http://dx.doi.org/10.1109/TNS.2007.909912}{\emph{IEEE
  Transactions on Nuclear Science} {\bfseries 54} (Dec, 2007) 2646--2652}.

\bibitem{SAULI20162}
F.~Sauli, \emph{The gas electron multiplier (gem): Operating principles and
  applications},
  \href{http://dx.doi.org/https://doi.org/10.1016/j.nima.2015.07.060}{\emph{Nuclear
  Instruments and Methods in Physics Research Section A: Accelerators,
  Spectrometers, Detectors and Associated Equipment} {\bfseries 805} (2016) 2
  -- 24}.

\bibitem{THGEM_operation_Ne_CH4}
M.~Cortesi, V.~Peskov, G.~Bartesaghi, J.~Miyamoto, S.~Cohen, R.~Chechik et~al.,
  \emph{Thgem operation in ne and ne/ch$_4$}, {\emph{Journal of
  Instrumentation} {\bfseries 4} (2009) P08001}.

\bibitem{cortesi1}
M.~Cortesi, J.~Yurkon and A.~Stolz, \emph{Operation of a thgem-based detector
  in low-pressure helium}, {\emph{Journal of Instrumentation} {\bfseries 10}
  (2015) P02012}.

\bibitem{cortesi2}
M.~Cortesi, J.~Yurkon, W.~Mittig, D.~Bazin, S.~Beceiro-Novo and A.~Stolz,
  \emph{Studies of thgem-based detector at low-pressure hydrogen/deuterium, for
  at-tpc applications}, {\emph{Journal of Instrumentation} {\bfseries 10}
  (2015) P09020}.

\bibitem{1748-0221-5-03-P03009}
M.~Alexeev, M.~Alfonsi, R.~Birsa, F.~Bradamante, A.~Bressan, M.~Chiosso et~al.,
  \emph{Development of thgem-based photon detectors for cherenkov imaging
  counters}, {\emph{Journal of Instrumentation} {\bfseries 5} (2010) P03009}.

\bibitem{2006physics/0606162}
R.~{Chechik}, M.~{Cortesi}, A.~{Breskin}, D.~{Vartsky}, D.~{Bar} and
  V.~{Dangendorf}, \emph{{Thick GEM-like (THGEM) detectors and their possible
  applications}}, {\emph{Proceedings of the SNIC Symposium on novel detectors
  April 3--6} (2006) },
  [\href{https://arxiv.org/abs/physics/0606162}{{\ttfamily physics/0606162}}].

\bibitem{Renous2017ogr}
D.~S. Renous, A.~Roy, A.~Breskin and S.~Bressler, \emph{Gain stabilization in
  micro pattern gaseous detectors: methodology and results}, {\emph{Journal of
  Instrumentation} {\bfseries 12} (2017) P09036}.

\bibitem{DallaTorre2015}
M.~Alexeev, R.~Birsa, F.~Bradamante, A.~Bressan, M.~B\"{u}chele, M.~Chiosso
  et~al., \emph{The gain in thick gem multipliers and its time-evolution},
  {\emph{Journal of Instrumentation} {\bfseries 10} (2015) P03026}.

\bibitem{Alfonsi20126}
M.~Alfonsi, G.~Croci, S.~D. Pinto, E.~Rocco, L.~Ropelewski, F.~Sauli et~al.,
  \emph{Simulation of the dielectric charging-up effect in a \mbox{GEM}
  detector}, \href{http://dx.doi.org/10.1016/j.nima.2011.12.059}{\emph{Nuclear
  Instruments and Methods in Physics Research Section A: Accelerators,
  Spectrometers, Detectors and Associated Equipment} {\bfseries 671} (2012) 6
  -- 9}.

\bibitem{Correia2014}
P.~M.~M. Correia, C.~A.~B. Oliveira, C.~D.~R. Azevedo, A.~L.~M. Silva,
  R.~Veenhof, M.~V. Nemallapudi et~al., \emph{A dynamic method for charging-up
  calculations: the case of gem}, {\emph{Journal of Instrumentation} {\bfseries
  9} (2014) P07025}.

\bibitem{2017arXiv170900095L}
{LUX Collaboration}, D.~S. {Akerib}, S.~{Alsum}, H.~M. {Ara{\'u}jo}, X.~{Bai},
  A.~J. {Bailey} et~al., \emph{{3D Modeling of Electric Fields in the LUX
  Detector}}, {\emph{ArXiv e-prints} (Aug., 2017) },
  [\href{https://arxiv.org/abs/1709.00095}{{\ttfamily 1709.00095}}].

\bibitem{garfieldpp}
``Garfield++.'' \url{http://cern.ch/garfieldpp/}.

\bibitem{github_chargingup}
P.~M.~M. Correia and M.~Pitt, ``Charging-up algorithm for garfield++.''
  \url{https://github.com/pmcorreia/Garfpp-chargingup.git}, 2017.

\bibitem{COliveira2012JINST}
C.~A.~B. Oliveira, P.~M.~M. Correia, H.~Schindler, A.~L. Ferreira, C.~M.~B.
  Monteiro, J.~M.~F. dos Santos et~al., \emph{Simulation of vuv
  electroluminescence in micropattern gaseous detectors: the case of \mbox{GEM}
  and \mbox{MHSP}}, {\emph{Journal of Instrumentation} {\bfseries 7} (2012)
  P09006}.

\bibitem{magboltz}
``Mabtoltz.'' \url{http://cern.ch/magboltz}.

\bibitem{Biagi1999234}
S.~Biagi, \emph{Monte carlo simulation of electron drift and diffusion in
  counting gases under the influence of electric and magnetic fields},
  \href{http://dx.doi.org/10.1016/S0168-9002(98)01233-9}{\emph{Nuclear
  Instruments and Methods in Physics Research Section A: Accelerators,
  Spectrometers, Detectors and Associated Equipment} {\bfseries 421} (1999) 234
  -- 240}.

\bibitem{1952PhRv.88.417J}
W.~P. {Jesse} and J.~{Sadauskis}, \emph{{Alpha-Particle Ionization in Mixtures
  of the Noble Gases}},
  \href{http://dx.doi.org/10.1103/PhysRev.88.417}{\emph{Physical Review}
  {\bfseries 88} (Oct., 1952) 417--418}.

\bibitem{AroucaTHGEMGAIN}
C.~Azevedo, P.~Correia, L.~Carramate, A.~Silva and J.~Veloso, \emph{Thgem gain
  calculations using garfield++: solving discrepancies between simulation and
  experimental data}, {\emph{Journal of Instrumentation} {\bfseries 11} (2016)
  P08018}.

\bibitem{Ozkan2010JINST}
{\"O}.~\c{S}ahin, I.~Tapan, E.~N. {\"O}zmutlu and R.~Veenhof, \emph{Penning
  transfer in argon-based gas mixtures}, {\emph{Journal of Instrumentation}
  {\bfseries 5} (2010) P05002}.

\bibitem{Sahin2014104}
{\"O}.~\c{S}ahin, T.~Z. Kowalski and R.~Veenhof, \emph{High-precision gas gain
  and energy transfer measurements in ar--co2 mixtures},
  \href{http://dx.doi.org/http://dx.doi.org/10.1016/j.nima.2014.09.061}{\emph{Nuclear
  Instruments and Methods in Physics Research Section A: Accelerators,
  Spectrometers, Detectors and Associated Equipment} {\bfseries 768} (2014) 104
  -- 111}.

\bibitem{Sauli:117989}
F.~Sauli, \emph{{Principles of operation of multiwire proportional and drift
  chambers}},  (Geneva), p.~92, CERN, CERN, 1977.

\end{thebibliography}\endgroup

\end{document}